%% file: main.tex
\pdfoutput=1
\documentclass[a4paper]{article}

\usepackage{INTERSPEECH2021}

\title{Graph-PIT: Generalized permutation invariant training for continuous separation of arbitrary numbers of speakers}  

\name{Thilo von Neumann$^1$, Keisuke Kinoshita$^2$, Christoph Boeddeker$^1$, Marc Delcroix$^2$, Reinhold Haeb-Umbach$^1$}
\address{
  $^1$ Paderborn University, Germany\\
  $^2$ NTT Corporation, Japan}
\email{\{vonneumann,boeddeker,haeb\}@nt.upb.de, \{keisuke.kinoshita,marc.delcroix\}@ieee.org}

\input{macros}
\input{glossaries}

\begin{document}

\maketitle
\begin{abstract}

Automatic transcription of meetings requires handling of overlapped speech, which calls for continuous speech separation (CSS) systems.
The uPIT criterion was proposed for utterance-level separation with neural networks and introduces the constraint that the total number of speakers must not exceed the number of output channels. 
When processing meeting-like data in a segment-wise manner, i.e., by separating overlapping segments independently and stitching adjacent segments to continuous output streams, this constraint has to be fulfilled for any segment.
In this contribution, we show that this constraint can be significantly relaxed.
We propose a novel graph-based PIT criterion, which casts the assignment of utterances to output channels in a graph coloring problem.
It only requires that the number of concurrently active speakers must not exceed the number of output channels. 
As a consequence, the system can process an arbitrary number of speakers and arbitrarily long segments and thus can handle more diverse scenarios.
Further, the stitching algorithm for obtaining a consistent output order in neighboring segments is of less importance and can even be eliminated completely, not the least reducing the computational effort.
Experiments on meeting-style WSJ data show improvements in recognition performance over using the uPIT criterion.

\end{abstract}

\def\nspk{K}
\def\nspki{k}
\def\nout{N}
\def\nouti{n}
\def\nutt{U}
\def\nutti{u}
\def\signal{\vect{s}}
\def\estimate{\hat{\signal}}
\def\coloringset{\mathcal{C}}
\def\graph{G}
\def\shift{l}
\def\meetingsignal{\vect{y}}
\def\historycontext{T_\text{h}}
\def\futurecontext{T_\text{f}}
\def\currentcontext{T_\text{c}}

\section{Introduction}

%
%
%

The automatic transcription of meetings has become a focus of research in recent years \cite{Neumann2019_AllneuralOnlineSource,Chen2020_ContinuousSpeechSeparation,Araki2018_ComparisonReferenceMicrophone,Carletta2006_AMIMeetingCorpus}.
Conventional speech analysis systems, such as speech recognition, are constructed for a single active speaker at a time \cite{Povey2011_KaldiSpeechRecognition,Watanabe2018_ESPnetEndtoEndSpeech}.
Since meeting recordings naturally contain overlapped speech, these systems cannot be applied directly, but require speech separation as pre-processing.

Many effective speech separation techniques have been proposed in the recent years based on neural networks, such as Deep Clustering \cite{Hershey2016_DeepClusteringDiscriminative} and models based on \gls{PIT} \cite{Yu2017_PermutationInvariantTraining,Kolbaek2017_MultitalkerSpeechSeparationa,Luo2018_TaSNetTimeDomainAudio,Luo2019_ConvTasNetSurpassingIdeal,Luo2020_DualPathRNNEfficient}.
Current state-of-the-art systems use the \gls{uPIT} \cite{Kolbaek2017_MultitalkerSpeechSeparationa} training scheme \cite{Luo2018_TaSNetTimeDomainAudio,Luo2019_ConvTasNetSurpassingIdeal,Luo2020_DualPathRNNEfficient}.
\gls{uPIT} training works by assigning each speaker to an output channel of a speech separation network such that the training loss is minimized.
This introduces the constraint that
the number of speakers $K$ must not exceed the number of output channels $N$ in the speech segment to be processed. 

As meetings can be of arbitrary length and can contain an arbitrary number of speakers, \gls{CSS} \cite{Chen2020_ContinuousSpeechSeparation}, i.e., handling of arbitrarily long audio streams, is required. 
\gls{CSS} can be realized by segmenting the input and processing the segments independently \cite{Yoshioka2018_RecognizingOverlappedSpeech,Chen2020_ContinuousSpeechSeparation}. 
Adjacent segments are aligned using a similarity measure in a so-called stitching process.
It was shown that the number of output channels of the source separator can be fixed to, say, $\nout=2$, although the total number of speakers $\nspk$ in a meeting may be much larger.
That is because, if we select sufficiently small segments, we can assume that normally the number of speakers that appear in such a short segment becomes equal to or smaller than $\nout$, regardless of $\nspk$.
Viewed differently, this means that the constraint of \gls{uPIT} effectively limits the segment size.
Even for the example of a relatively short segment size of \SI{2.4}{\second}, more than \SI{22}{\percent} of the segments contain more than two speakers in the CHiME-5 \cite{Barker2018_FifthCHiMESpeech} evaluation dataset.
In addition, the constraint of \gls{uPIT} cannot be fulfilled when applied to or trained on full meetings.


In this contribution, we propose a generalization of \gls{uPIT}, which relaxes of the above constraint ($K \le N$).
The generalization is achieved by incorporating the idea that different speakers can be put on the same output channel as long as they never overlap.
We reformulate the problem of assigning utterances to output channels as a graph coloring problem, hence the name \gls{mPIT}.
With \gls{mPIT}, we only need to ask for  the number of \textit{concurrently} active speakers, i.e., speakers speaking at the same time, to not exceed the number of output channels. 
This constraint is far more realistic and easier to satisfy compared to the original constraint imposed by \gls{uPIT}.
Looking again at the CHiME-5 evaluation dataset, this constraint is only violated \SI{9}{\percent} of the time compared to \SI{22}{\percent} for the \gls{uPIT} constraint in case of $\nout=2$.
In a stitching-based \gls{CSS} scenario, the proposed \gls{mPIT} criterion allows for arbitrarily long  segments and arbitrarily many speakers in a segment as long as no more than $N$ speakers speak at a time.
Moreover, in a general CSS scenario without segmentation and stitching, \gls{mPIT} theoretically allows modeling the entire meeting with a separator that can utilize any contextual information.


In our experiments, we show the effectiveness of the proposed \gls{mPIT} loss on simulated meetings based on WSJ data.
We can increase the segment size for stitching significantly and show that stitching is not even necessary for two-minute long meeting-like data.

\section{Continuous Speech Separation}

\gls{CSS} \cite{Chen2020_ContinuousSpeechSeparation} describes the task of separating an input audio signal $\meetingsignal$ into one or multiple overlap-free signals $\estimate_\nouti, \nouti=1,...,\nout$.
We model a meeting $\meetingsignal$ as the sum of $\nutt$ utterance signals produced by $\nspk$ different speakers, where utterances of different speakers may overlap:
\begin{equation}
    \meetingsignal=\sum_{\nutti=1}^\nutt \signal_\nutti.
\end{equation}
The signal $\signal_\nutti$ is the $\nutti$-th utterance, shifted and zero-padded to the length of the full meeting.

Attempts to solve the \gls{CSS} problem led to multiple different approaches, e.g., yielding one signal per utterance ($\nout>\nspk$) \cite{Kanda2019_SimultaneousSpeechRecognition,Watanabe2020_CHiME6ChallengeTackling} or producing one continuous stream per speaker \cite{Hori2012_LowLatencyRealTimeMeeting,Neumann2019_AllneuralOnlineSource} ($\nspk=\nout$).
Here, we concentrate on the ideas proposed in \cite{Yoshioka2018_RecognizingOverlappedSpeech}.





The approach from \cite{Yoshioka2018_RecognizingOverlappedSpeech} is based on the idea that the number of concurrently speaking speakers is usually much smaller than total number of speakers in a meeting, i.e., less output channels than speakers are required ($\nout\leq\nspk$).
It thus segments the meeting signal $\meetingsignal$ into overlapping segments, separates the speech in each segment independently into $\nout$ signals, and uses a stitching mechanism (\cref{sec:stitching}) to align the output streams across segments.
The neural-network-based separator is trained with \gls{uPIT} (\cref{sec:uPIT}), which imposes the constraint that the number of speakers in a segment must not exceed the number of output channels, $\nout$.

We propose to replace \gls{uPIT} with \gls{mPIT} (\cref{sec:mPIT}) to relax this constraint, only requiring that the number of simultaneously speaking speakers is smaller than the number of output channels.
In the remainder of this section, we look at a segment of a meeting, i.e., $\nspk$ represents the number of speakers in the segment and $\signal_\nutti$ is scoped to this segment, to simplify explanations.
We assume a time-domain source separator with $\nout$ output channels.




\subsection{Utterance-level PIT}
\label{sec:uPIT}

\def\upitmapping{\pi^{\text{(\acs{uPIT})}}}
\def\permset{\mathcal{P}}
\def\spksignal{\signal^\text{(spk)}}

The traditional \gls{uPIT} \cite{Kolbaek2017_MultitalkerSpeechSeparationa}  assigns each speaker to an output channel, i.e., $\nout=\nspk$.\footnote{This can be relaxed to $\nspk<\nout$ by adding silent target signals $\signal_i=0$ for ${K < i \leq N}$.}
During training, the permutation problem between targets and estimated audio streams is solved by finding the bijective mapping $\upitmapping:\{1,...,\nspk\}\rightarrow\{1,...,\nout\}$ between speakers and output channels that minimizes the loss:
\begin{equation}
    \loss{uPIT} = \min_{\upitmapping \in \permset_\nout} \sum_{\nspki=1}^{\nspk} \L(\spksignal_\nspki, \estimate_{\upitmapping(\nspki)}).
\end{equation}
The loss function $\L$ is a signal-level loss function, $\permset_\nout$ is the set of all permutations of length $\nout$ and $\spksignal_\nspki$ is the sum of all utterances of speaker $\nspki$.

\subsection{Graph-based meeting-level PIT}
\label{sec:mPIT}
\def\mpitmapping{\pi^{\text{(\acs{mPIT})}}}
\def\helper{\tilde{\signal}}
\def\targetset{\mathcal{U}}

\begin{figure}
    \centering
    \input{tikz/graph-example.tikz.tex}
    \caption{
        Example of processing a three-speaker scenario using \acs{mPIT} with a two-output separator.
        Each box represents one utterance.
        \emph{Top}: Utterances in the meeting and the colored overlap graph $G$.  \gls{mPIT} is equivalent to \gls{uPIT} for an activity pattern as marked with (a).
        \emph{Bottom}: A possible mapping of utterances to output channels.}
    \label{fig:graph}
\end{figure}
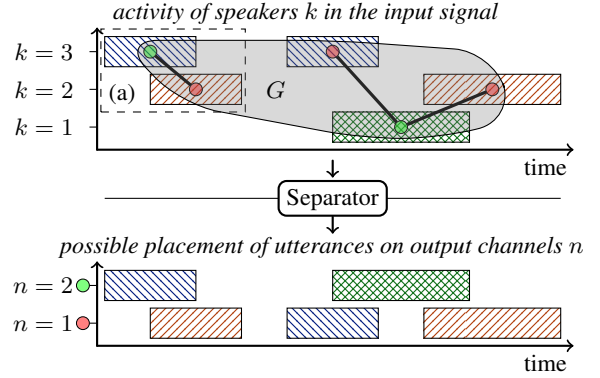

If we relax the constraint $\nspk=\nout$ of \gls{uPIT} to $\nout\leq\nspk$, i.e., an output channel is no longer bound to a speaker, there is no bijective mapping between output channels and speakers.
We can, however, find an, in general, non-bijective mapping ${\pi: \{1,...,\nutt\} \rightarrow \{1,...,\nout\}}$ of target utterances to output channels so that overlapped utterances are separated.

Finding such a mapping is equivalent to a graph coloring problem \cite{Hopcroft2006_AutomataTheoryLanguages}; if each utterance is modelled as a vertex and edges are drawn between utterances that overlap, the set of all proper $\nout$-vertex-colorings of this graph is equal to the set of mappings from utterances to output channels.
A $\nout$-vertex-coloring assigns each vertex a color from a set of $\nout$ colors so that connected vertices should be colored differently.

\def\vertexset{V}
\def\edgeset{E}
%
This graph $\graph=(V,E)$ is undirected and defined with
\begin{equation}
    \begin{aligned}
        \vertexset & = \left\{1,\dots,\nutt\right\},\\
        \edgeset & = \left\{\left\{u,v\right\} : \forall u, v \in \vertexset, u \neq v \text{ if } \signal_u \text{ and } \signal_v \text{ overlap} \right\},
    \end{aligned}
\end{equation}
where $\vertexset$ are the vertices / utterances and $\edgeset$ the edges between overlapping utterances.
An example of such a graph with two connected components is shown in the top part of \cref{fig:graph}.
A proper $\nout$-vertex-coloring of $G$ is defined as a mapping 
\begin{equation}
    \begin{aligned}
        \mpitmapping: \vertexset\rightarrow\{1,...,N\}, \text{ such that} \\
        \mpitmapping(u)\neq\mpitmapping(v)\quad\forall\{u, v\}\in\edgeset.
    \end{aligned}
\end{equation}
Note that $\mpitmapping$ does not have to be surjective, i.e., the mapping is not required to use all output channels.
Let $\coloringset_{\graph,\nout}$ be the set of all proper $\nout$-vertex-colorings of $G$.
%
Then, we formulate the \gls{mPIT}\footnote{Code is available at \url{https://github.com/fgnt/graph_pit}.} objective as
%


\begin{equation}
    \loss{\acs{mPIT}} = \min_{\mpitmapping\in\coloringset_{\graph,\nout}} \sum_{\nouti=1}^{\nout} \L\left(\helper_{\mpitmapping,\nouti}, \estimate_\nouti\right),
\label{eq:mPIT}
\end{equation}
where we construct an intermediate target signal
\begin{equation}
    \helper_{\pi,\nouti} = \sum_{\nutti = 1}^{\nutt} \begin{cases}
         \signal_\nutti, & \text{if } \pi(\nutti) = \nouti, \\
        \vect{0},& \text{otherwise.} \\
    \end{cases}
\end{equation}
The set $\coloringset_{\graph,\nout}$ is computed by enumerating all graph colorings.


An example graph $G$ together with a possible mapping of target utterances to output channels, i.e., a coloring, is drawn for a fictive speaker activity pattern in \cref{fig:graph}.
\gls{mPIT} is equivalent to \gls{uPIT} for a connected graph with two speakers only; an example of this is marked with (a).
If the graph consists of more than one connected component (assuming $\nspk\leq\nout$) \gls{uPIT} places all utterances of a single speaker on the same output channel, which enforces the model to use global information.
\gls{mPIT} gives more freedom to the placement as individual connected components are treated separately.
If there are more than $\nout$ speakers in a connected component (as in the second connected component of $G$ in \cref{fig:graph}), \gls{mPIT} can provide a solution for the assignment problem while \gls{uPIT} cannot.

\subsection{Complexity}
The vertex coloring problem is in general NP-hard \cite{Hopcroft2006_AutomataTheoryLanguages}.
While \gls{uPIT} computes $\nout!$ permutations, the number of valid colorings for a graph is bounded by $\nout^\nutt$ (in the extreme case of no edges), which can easily exceed $\nout!$ if $\nutt >> \nout$.
However, $N$ and $U$ are typically small in a segment used for training and it is unlikely that the graph has no edges so that the computational overhead is negligible in comparison to the neural network.




\subsection{Thresholded SDR for varying numbers of speakers}

For this paper, we use a time-domain separation model as a basis of investigation. 
However, the \gls{SDR}-based losses, which have shown to be effective for training such models  \cite{Luo2019_ConvTasNetSurpassingIdeal,Luo2020_DualPathRNNEfficient}, are problematic when used in training with meeting-like data containing significant amounts of silence.
Specifically, these losses have the problem that their value is unbounded so that easy examples can dominate the training and that they are undefined for silent target signals, i.e., $\signal = \vect{0}$.
Silent targets are required for training an $N$-output separator with less than $N$ target utterances, e.g., when the mapping $\mpitmapping$ is not surjective in \cref{eq:mPIT}.
The first problem can be solved by the \gls{tSDR} loss \cite{Wisdom2020_UnsupervisedSpeechSeparation}\footnote{\acs{tSDR} is called \enquote{thresholded signal-to-noise ratio} in \cite{Wisdom2020_UnsupervisedSpeechSeparation}, but it measures distortions rather than noise in this case.} which limits the value of the loss function at a soft maximum.
The second problem can be solved by adding a small constant $\varepsilon$, ending up with the $\varepsilon$-\gls{tSDR} loss
%
\begin{equation}
\begin{aligned}
    \loss{$\varepsilon$-tSDR} (\vect{\signal}, \vect{\estimate}) &= -10 \log_{10} \frac{|\vect{\signal}|^2 + \varepsilon}{|\vect{\signal} - \vect{\estimate}|^2 + \tau(|\vect{\signal}|^2 + \varepsilon)}\\
    &\geq -\text{SDR}_\text{max},
\end{aligned}
\end{equation}
where $\tau=10^{-\text{SDR}_\text{max} / 10}$ introduces soft threshold at $\text{SDR}_\text{max}$.
The constant $\varepsilon$ makes sure that the loss is defined even if $\signal=\vect{0}$.

\section{Experiments}

In this section, we investigate separation networks trained with \gls{uPIT} and \gls{mPIT} as meeting-level separators and segment-level separators in a stitching-based system.
We expect that \gls{uPIT} works only in the stitching-based case, while \gls{mPIT} works for batch-processing as well as segment-level separation.

\subsection{Stitching for CSS}
\label{sec:stitching}

For the stitching-based system, we use a processing scheme similar to \cite{Yoshioka2018_MultiMicrophoneNeuralSpeech,Chen2020_ContinuousSpeechSeparation}.
The incoming audio is segmented into overlapping segments.
Each segment is processed by the separation network independently.
A segment consists of three sub-segments representing a history, current and future context with lengths of $\historycontext$, $\currentcontext$ and $\futurecontext$ seconds, respectively.
Segments are shifted by $\currentcontext$.
The current context is used for reconstruction while history and future contexts improve the separation quality at segment edges and are used to align output streams based on squared differences between overlapping sub-segments.

\subsection{Data}
For our experiments, we generate artificial meetings based on the WSJ \cite{Garofalo2007_CsriWsj0Complete} corpus.
We randomly sample five to eight speakers for each meeting so that the distribution of speakers is uniform across all meetings.
We also sample an overlap ratio between $0.2$ and $0.4$ for each meeting.
Then, we uniformly 
select utterances of the sampled speakers and sample a start times so that the overlap ratio is roughly fulfilled and speakers are active for roughly the same amount of time.
Short silence is sampled between two utterances with a probability of \SI{10}{\percent}.
All utterances of a speaker are scaled with the same logarithmic weight uniformly drawn from \SIrange{0}{5}{\decibel}.
The meetings are corrupted by \SIrange{20}{30}{\decibel} of simulated white microphone noise.

The train and validation sets are based on \emph{train\_si284} and \emph{cv\_dev93}
, respectively.
The evaluation set is based on \emph{test\_eval92} with a total length of about \SI{1}{\hour}.
The meetings are about \SI{120}{\second} long which matches the length of segments used for meeting-wise evaluation in libri-CSS \cite{Chen2020_ContinuousSpeechSeparation}.
We intentionally stick to clean reverberation-free meetings based on WSJ instead of the libri-CSS database \cite{Chen2020_ContinuousSpeechSeparation} for a proof-of-concept of the \gls{mPIT} objective.
Libri-CSS is based on Librispeech \cite{Panayotov2015_LibrispeechASRCorpus} which contains utterances with large portions of silence. 
This makes it unclear how to define utterance boundaries.
We use a sample rate of \SI{8}{\kilo\hertz} to speed up the experiments.

\subsection{Metrics}

For evaluation, we use metrics computed with an oracle diarization system.
We perform separation in a continuous manner, i.e., we feed whole meetings into the separator, but compute the metrics utterance-wise by using oracle utterance boundaries.
This scheme gives an upper bound on the performance.
We compute the \gls{WER} and \gls{SDR} \cite{Vincent2006_PerformanceMeasurementBlind} improvement (SDRi) compared to the unprocessed meeting.

We use an End-to-End ASR model from ESPnet \cite{Watanabe2018_ESPnetEndtoEndSpeech} trained on clean WSJ data re-sampled to \SI{8}{\kilo\hertz} to compute the \gls{WER}.
The model achieves a \gls{WER} of \SI{5.6}{\percent} on the clean eval92 set of WSJ.
We use the mir\_eval toolbox \cite{Raffel2014_MirEvalTransparent} to obtain the \gls{SDR}i.

\subsection{Model architecture and training scheme}
As the separation model, we use a \gls{DPRNN}-based \gls{TasNet} \cite{Luo2020_DualPathRNNEfficient} with two output channels, so we fix $\nout=2$.
To keep the computational cost low, we work with a smaller model with three blocks instead of six.
The remaining configuration is the same as \cite{Luo2020_DualPathRNNEfficient}: We set the number of filters in the encoder and decoder to 64, the number of hidden units to 128 in each direction, and the chunk length to 100.
We use the $\varepsilon$-tSDR loss for all models with $\text{SNR}_\text{max}=\text{\SI{20}{\decibel}}$ and $\varepsilon=10^{-6}$.
Our architecture achieves \SI{15}{\decibel} \gls{SDR} gain on WSJ0-2mix \cite{Hershey2016_DeepClusteringDiscriminative}.

\subsubsection{Baseline uPIT model}
\noindent
We train the baseline system similar to \cite{Yoshioka2018_RecognizingOverlappedSpeech} but with \gls{DPRNN}-\gls{TasNet} with \gls{uPIT} on segments of meeting-like data, discarding any segments containing more than two speakers to match the evaluation data as closely as possible.
We adjust the batch size so that all models see the same amount of \SI{32}{\second} of data per batch when training with different segment lengths $T_\text{tr}$.

\subsubsection{Graph-PIT model}
\noindent
We train our proposed model with the following schedule:
We train the model on segments of meeting-like data with \gls{uPIT} like the baseline model which eases convergence at the beginning of training.
Then, we re-train the pre-trained model with the \gls{mPIT} loss on the full training data including segments with more than two speakers.
We reduce the amount of single-speaker examples during re-training to obtain a significant amount of examples with more than two speakers.







\subsection{Meeting-wise evaluation}

\begin{table}[t]
    \centering
    \caption{
        Performance on meetings of about \SI{120}{\second} length with five to eight speakers. 
        Only the best stitcher configurations are shown. $T_\text{tr}$ is the length of training segments in seconds. 
        Experiments without stitching are marked with \xmark.
        Best results are \textbf{bold} and the best results without stitching are \underline{underlined}.
    }
    \label{tbl:long}
    \setlength{\tabcolsep}{1.5pt}
    \input{tables/long.tex}
\end{table}

\noindent
\cref{tbl:long} shows the performance of different separation models on meetings with a length of about \SI{120}{s}.
We show results obtained with meeting-level batch processing as well as with stitching-based \gls{CSS}. 
For stitching-based \gls{CSS}, we only show the results for the best stitcher configuration for each model, determined by keeping $\historycontext$ and $\futurecontext$ constant at \SI{0.4}{\second} and sweeping $\currentcontext$ from \SIrange{1}{14}{\second} on the development set.
We observed that the stitching process works reasonably well for $\historycontext=\futurecontext=\text{\SI{1}{\second}}$.
A comparison of different stitcher configurations is given in \cref{sec:eval:stitching}.
We only report results for \gls{mPIT} where we expect differences to \gls{uPIT}; for small segment sizes that are usually used for training of \gls{uPIT}, the probability of seeing more than two speakers in a training and test segment is small.

\cref{tbl:long} shows the WER grouped by number of overlapping speakers per utterance, and the \gls{WER} and \gls{SDR}i for the whole meetings.
When looking at the overall performance, we see a slight improvement for training closer to the evaluation condition with \gls{mPIT} over \gls{uPIT} when they are used in stitching based CSS framework.
Models trained with \gls{mPIT} generalize to processing the whole meeting of \SI{120}{\second} length at once without using a stitcher (\SI{13.0}{\percent} \gls{WER}) when trained with long enough segments while \gls{uPIT} does not handle this case well (\SI{18.4}{\percent} \gls{WER}).
That is because \gls{uPIT} is, by its constraint, not built to handle more than two speakers as it is the case for full meetings, while \gls{mPIT} trains for this scenario.
Longer training segments match the evaluation scenario better.

When looking at the performance with respect to the number of speakers per utterance, we observe that the largest improvement of \gls{mPIT} over \gls{uPIT} comes from utterances that overlap with two other speakers (i.e., num. spk. = 3).
This is the scenario that \gls{mPIT} is explicitly trained for but \gls{uPIT} is not.
The improvement for non-overlapped utterances (i.e., num. spk. = 1) over no separation comes from suppressing the microphone noise.
Shorter training segments are better if the model is applied with stitching, while longer training segments are better when evaluated without stitching, for both objectives.


\subsection{The effect of stitching}
\label{sec:eval:stitching}

\begin{figure}
    \centering
    \input{tikz/plot-stitcher.tikz.tex}
    \vspace{-.5cm}
    \caption{\emph{Top}: \acs{WER} plotted over the segment size for stitching. \emph{Bottom}: Distribution of the number of speakers in a segment. The red line represents the amount of segments that fulfill the constraint of \acs{uPIT}.}
    \label{fig:stitcher}
    \vspace{-2mm}
\end{figure}
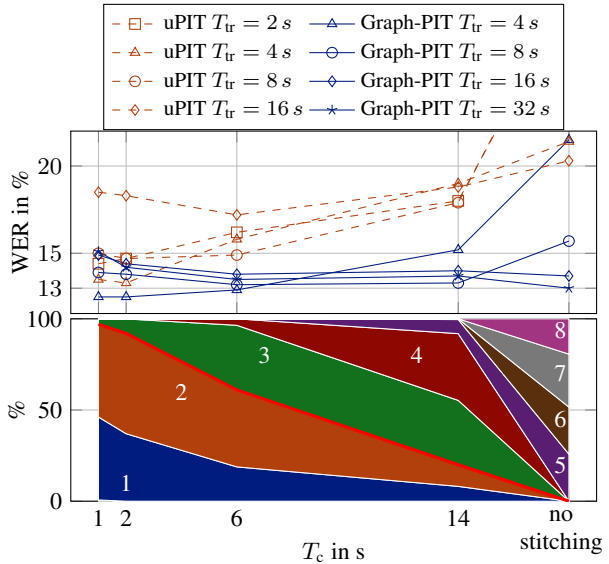


%

\cref{fig:stitcher} shows the effect of stitching on the performance in \gls{WER}.
The bottom plot shows the distribution of the number of speakers in a segment.
The red line represents the amount of segments that fulfill the number-of-speaker constraint ($K\leq N$) of \gls{uPIT}.
As expected, the \gls{uPIT} model gets degraded with larger segments where its constraint is violated.
The performance of the \gls{mPIT} model, on the other hand, improves with increasing segment sizes and even generalizes to processing the whole meeting at once.
\gls{mPIT} shows a better \gls{WER} than \gls{uPIT} for all scenarios.

One advantage of larger segments for stitching is the reduced computational overhead that scales linearly with the segment overlap ratio.
We can reduce the segment overlap ratio significantly from \SI{200}{\percent} to \SI{14}{\percent} by increasing the segment size from \SI{3}{\second} to \SI{16}{\second} and can eliminate the stitching process completely in our experiments on meetings with a length of \SI{120}{\second}.


It is interesting to see that the \gls{uPIT} model, even though trained only on data with up to two speakers, does not break completely when it sees data violating its constraint. 
This can be seen in \cref{fig:stitcher} and \cref{tbl:long}, i.e., \SI{49.1}{\percent} \gls{WER} with no processing and \SI{18.4}{\percent} with batch processing.
However, employing \gls{mPIT} during training drastically improves the performance in these cases.




\section{Conclusions}

In this paper, we proposed a generalization of \gls{uPIT} for \gls{CSS}-style processing of long recordings, called \gls{mPIT}.
\gls{mPIT} relaxes the constraint of having less speakers than output channels in one segment to having less concurrent speakers (i.e., speakers active in one sample) than output channels.
It thus enables processing of more diverse meetings and the use of much larger segments while reducing the computational overhead introduced by stitching.
We showed that the \gls{mPIT} objective can be used to construct separation networks that do not require stitching at all.
In future work, we plan to test \gls{mPIT} on more realistic data like libri-CSS \cite{Chen2020_ContinuousSpeechSeparation} or real recordings.

\section{Acknowledgements}
Computational resources were provided by the Paderborn Center for Parallel Computing.

\bibliographystyle{IEEEtran}

\balance
\bibliography{references}


\end{document}

%% file: macros.tex
\usepackage{amsmath}
\usepackage{amssymb}
\usepackage{bbm}
\usepackage{physics}
\usepackage{siunitx}
\usepackage{trfsigns}
\usepackage{pifont}
\usepackage{etoolbox}
\usepackage{multirow}
\usepackage{csquotes}
\usepackage[shortlabels]{enumitem}  
\usepackage{cite}   
\usepackage{hyperref}
\usepackage[capitalize]{cleveref}
\usepackage{tikz}
\usepackage{balance} 

\usepackage{url}

\newcommand{\vect}[1]{\ensuremath{\boldsymbol{\mathbf{#1}}}}
\def\L{\ensuremath{\mathcal{L}}}

\newcommand{\xmark}{{--}}

\newcolumntype{H}{>{\setbox0=\hbox\bgroup}c<{\egroup}@{}}   

\newcommand{\tab}[2][c]{\begin{tabular}{@{}#1@{}}#2\end{tabular}}   

\newcommand*{\thl}{\fontseries{b}\selectfont}
\robustify\thl
\robustify\fontseries

\newcommand{\loss}[1]{\mathcal{L}^{\text{(#1)}}}

\usepackage{pgfplots}
\pgfplotsset{compat=1.16}
\usepgfplotslibrary{groupplots}
\usetikzlibrary{positioning,arrows,matrix,fit,calc,patterns,chains,scopes,shapes.multipart,decorations,decorations.markings,backgrounds}

\definecolor{palette-1}{HTML}{001C7F}    
\definecolor{palette-2}{HTML}{B1400D}    
\definecolor{palette-3}{HTML}{12711C}    
\definecolor{palette-4}{HTML}{8C0800}    
\definecolor{palette-5}{HTML}{591E71}    
\definecolor{palette-6}{HTML}{592F0D}    
\definecolor{palette-7}{HTML}{A23582}    
\definecolor{palette-8}{HTML}{3C3C3C}    
\definecolor{palette-9}{HTML}{B8850A}    
\definecolor{palette-10}{HTML}{006374}      

\tikzset{
    line/.style={draw,black,thick,rounded corners=1mm,line cap=round},
    noshortarrow/.style={line,->},
    arrow/.style={noshortarrow,shorten >=.3mm},
    doublearrow/.style={arrow,<->, shorten <=.3mm},
    box/.style={draw,black,thick,minimum height=3em,text height=1.5ex,text depth=0.25ex,rounded corners=3,fill=white},
    nopadding/.style={minimum height=0,inner sep=1mm},
    signalbox/.style={draw,black,thin,rounded corners=1mm,minimum width=7mm, minimum height=4mm,inner sep=0},
    pbox/.style={box,fill=black!10},
    backgroundbox/.style={inner xsep=3mm, inner ysep=1mm, draw, dashed, rounded corners,fill=orange!10},
    branch/.style={inner sep=0.3mm,circle,fill=black},
    operator/.style={draw,circle,black,rounded corners,inner sep=0,fill=white},
    vertex/.style={draw,ultra thin,circle,black,rounded corners,inner sep=0.6mm,fill=gray,fill opacity=0.5},
    edge/.style={line,very thick,line cap=butt},
    pattern1/.style={pattern=north west lines,pattern color=palette-1},
    pattern2/.style={pattern=north east lines,pattern color=palette-2},
    pattern3/.style={pattern=crosshatch,pattern color=palette-3},
    buswidth/.style={path picture={\draw[black,-] (path picture bounding box.south west) -- (path picture bounding box.north east);}}
}

%% file: glossaries.tex
\usepackage[acronym,shortcuts]{glossaries}
\makeglossaries
\glsdisablehyper
\newacronym{ASR}{ASR}{automatic speech recognition}
\newacronym{DPRNN}{DPRNN}{Dual-Path Recurrent Neural Network}
\newacronym{TasNet}{TasNet}{Time-domain Audio Separation Network}
\newacronym{WER}{WER}{Word Error Rate}
\newacronym{CER}{CER}{Character Error Rate}
\newacronym{PIT}{PIT}{Permutation Invariant Training}
\newacronym{uPIT}{uPIT}{Utterance-level \gls{PIT}}
\newacronym{mPIT}{Graph-PIT}{Graph-based Permutation Invariant Training}
\newacronym{BLSTM}{BLSTM}{Bi-directional Long Short-Term Memory Network}
\newacronym{CSS}{CSS}{Continuous Speech Separation}
\newacronym{SDR}{SDR}{Signal-to-Distortion Ratio}
\newacronym{tSDR}{tSDR}{Thresholded \gls{SDR}}
\newacronym{VAD}{VAD}{Voice Activity Detection}
\newacronym{NN}{NN}{Neural Network}

%% file: tikz/graph-example.tikz.tex
\begin{tikzpicture}
    \def\t{0.2}
    \newcommand{\rect}[3]{($(c#2|-r#1) + (0, \t)$) rectangle ($(c#3|-r#1) + (0, -\t)$)}

    \path (0, 0) --coordinate[pos=0](c0) coordinate[pos=1/10](c1) coordinate[pos=2/10](c2) coordinate[pos=3/10](c3) coordinate[pos=4/10](c4) coordinate[pos=5/10](c5) coordinate[pos=6/10](c6) coordinate[pos=7/10](c7) coordinate[pos=8/10](c8) coordinate[pos=1](c9) (6, 0);

    \path (0, 0) --coordinate[pos=0](r0) coordinate[pos=1/2](r1) coordinate[pos=1](r2) (0, -1);
    \node[anchor=south west] at ($(r0-|c0) + (-0,0.25)$) {\it{activity of speakers $\nspki$ in the input signal}};
    \draw[black] ($(r0-|c0) + (-0.1,0)$) -- ++(-0.1, 0) node[left,xshift=-1mm]{$\nspki=3$};
    \draw[black] ($(r1-|c0) + (-0.1,0)$) -- ++(-0.1, 0) node[left,xshift=-1mm]{$\nspki=2$};
    \draw[black] ($(r2-|c0) + (-0.1,0)$) -- ++(-0.1, 0) node[left,xshift=-1mm]{$\nspki=1$};

    \draw[pattern1] \rect{0}{0}{2} node[pos=0.5,vertex,fill=green](v11){};
    \draw[pattern1] \rect{0}{4}{6} node[pos=0.5,vertex,fill=red](v12){};

    \draw[pattern2] \rect{1}{1}{3} node[pos=0.5,vertex,fill=red](v21){};
    \draw[pattern2] \rect{1}{7}{9} node[pos=0.5,vertex,fill=red](v22){};
    
    \draw[pattern3] \rect{2}{5}{8} node[pos=0.5,vertex,fill=green](v31){};

    \node[draw,dashed,fit={(r0-|c0)(r1-|c3)},inner ysep=3mm, inner xsep=0.5mm](box){};
    \node[anchor=south west,inner sep=1pt,xshift=2pt,yshift=2pt] at (box.south west) {(a)};

    \draw[edge] (v11) -- (v21);
    \draw[edge] (v12) -- (v31) -- (v22);

    \draw[fill=gray, fill opacity=0.3, rounded corners=0.5cm] ($(v11) + (-0.4,0.15)$) -- ($(v12) + (0, 0.15)$) -- ($(v22) + (0, 0.5)$) --  ($(v22) + (0.3, -0.4)$) -- ($(v31) + (0.5, -0.15)$) -- ($(v31) + (-0.5, -0.15)$) -- ($(v21) + (-0.2, -0.2)$) -- cycle;
    \path (v11) -- (v31) node [pos=0.5,text=black]{$G$};

    \path (r2) -- ++(0, -2.1) -- coordinate[pos=0] (r3) coordinate[pos=1] (r4) ++(0, -0.5);
    \node[anchor=south west] at ($(r3-|c0) + (-0.7,0.25)$) {\it{possible placement of utterances on output channels $\nouti$}};
    \draw[black] ($(r3-|c0) + (-0.1,0)$) -- ++(-0.1, 0) node[left,xshift=-1mm]{$\nouti=2$}node[left,vertex,fill=green]{};
    \draw[black] ($(r4-|c0) + (-0.1,0)$) -- ++(-0.1, 0) node[left,xshift=-1mm]{$\nouti=1$}node[left,vertex,fill=red]{};

    \draw[pattern1] \rect{3}{0}{2};
    \draw[pattern1] \rect{4}{4}{6};
    \draw[pattern2] \rect{4}{1}{3};
    \draw[pattern2] \rect{4}{7}{9};
    \draw[pattern3] \rect{3}{5}{8};
    
    \coordinate (sepy) at ($(r2)!0.45!(r3)$);
    \draw (sepy-|c0) -- (sepy-|c9) node[box,pos=0.5,minimum height=0](sep){Separator};
    \draw[arrow] ($(sep) + (0,0.5)$) -- (sep);
    \draw[arrow] (sep) -- ($(sep) - (0,0.5)$);
    
    \draw[arrow] ($(r2-|c0) + (-0.1,-0.3)$) -- ($(r2-|c9) + (0.2,-0.3)$)  node[pos=1, below, anchor=north east]{time};
    \draw[arrow] ($(r2-|c0) + (-0.1,-0.3)$) -- ($(r0-|c0) + (-0.1,0.35)$);
    
    \draw[arrow] ($(r4-|c0) + (-0.1,-0.3)$) -- ($(r4-|c9) + (0.2,-0.3)$)  node[pos=1, below, anchor=north east]{time};
    \draw[arrow] ($(r4-|c0) + (-0.1,-0.3)$) -- ($(r3-|c0) + (-0.1,0.35)$);

\end{tikzpicture}

%% file: tables/long.tex

\robustify\underline

\newcommand{\Uline}[1]{#1\llap{\underline{#1}}}

\begin{tabular}{lS[table-format=2.0]ccSS[table-format=2.1]S[table-format=2.1]cS[table-format=2.1]S[table-format=2.1]}

\toprule
Model & {$T_\text{tr}$} & Stitching &  & \multicolumn{3}{c}{WER for num. spk} && \multicolumn{2}{c}{total}\\
\cmidrule{5-7}\cmidrule{9-10}
&&$\historycontext$+$\currentcontext$+$\futurecontext$&& {1} & {2} & {3} && {SDRi} & {WER} \\
\midrule
No sep. & & & &  
   10.9 & 45.9 & 67.4 && 0.0 & 49.1 \\
\midrule
\acs{uPIT}  & 2  & 1+0.4+1 &&  7.5 & 14.1 & 16.4 && 11.8 & 14.1 \\
            &    & \xmark && 10.1 & 28.3 & 38.8 && 4.6 & 30.0 \\
            & 4  & 1+2+1  && 8.1 & 12.6 & 16.2 && 12.1 & 13.3 \\
            &    & \xmark && 9.7 & 27.7 & 41.6 && 3.9 & 30.3 \\
            & 8  & 1+2+1  && 8.8 & 13.7 & 18.1 && 11.5 & 14.6 \\
            &    & \xmark && 8.0 & 17.7 & 26.9 && 8.9 & 19.7 \\
            & 16 & 1+6+1  && 8.7 & 16.4 & 21.5 && 10.1 & 17.2 \\
            &    & \xmark && 8.0 & 17.3 & 23.6 && 9.3 & 18.4 \\
\midrule
\acs{mPIT}  & 4  & 1+1+1 && 7.5 & \thl 12.0 & \thl 15.0 && \thl 12.6 & \thl 12.5 \\
            &    & \xmark && 7.2 & 20.7 & 27.8 && 8.1 & 21.5 \\
            & 8  & 1+6+1 && 7.7 & 12.8 & 16.1 && 12.3 & 13.2 \\
            &    & \xmark && 8.2 & 14.8 & 19.6 && 10.9 & 15.6 \\
            & 16 & 1+6+1 && 9.0 & 13.2 & 16.4 && 11.9 & 13.8 \\
            &    & \xmark && 7.1 & 13.2 &  16.7 && 11.9 & 13.7 \\
            & 32 & 1+6+1 && 7.8 & 12.5 & 16.9 && 12.0 & 13.5 \\
            &    & \xmark && \Uline{\thl 7.0} & \Uline{12.2} & \Uline{16.2} && \Uline{12.1} & \Uline{13.0} \\
\bottomrule
\end{tabular}

%% file: tikz/plot-stitcher.tikz.tex
\begin{tikzpicture}
    \def\mystrut{\vphantom{hg}}

    \pgfplotsset{
        mPIT/.style={
                color=palette-1,
                mark size=2pt,
            },
        uPIT/.style={
                color=palette-2,
                dashed,
                mark options={solid},
                mark size=2pt,
            }
    }

    \begin{groupplot}[
            group style={
                    group size=1 by 2,
                    vertical sep=2pt,
                    x descriptions at=edge bottom,
                },
            grid=both,
            height=4cm,
            width=8.5cm,
            legend,
            legend style={font=\footnotesize},
            y label style={yshift=-0.1cm},
            xtick={1,2,6,14,18},
            xmin=0, xmax=19,
        ]
        \nextgroupplot[ylabel=\acs{WER} in \%,
            legend to name={legend},
            legend columns=2,
            legend style={
                legend cell align=left,
            },
            ytick={13,15,20},
            ymax=22,
        ]

        \addplot[uPIT,mark=square] table[x=stitcher_size, y=upit2s] {data/stitching-mpit-1sh.dat};
        \addlegendentry{\acs{uPIT} $T_\text{tr}=2\,s$}
        \addplot[mPIT,mark=triangle] table[x=stitcher_size, y=mpit4s] {data/stitching-mpit-1sh.dat};
        \addlegendentry{\acs{mPIT} $T_\text{tr}=4\,s$}
        \addplot[uPIT,mark=triangle] table[x=stitcher_size, y=upit4s] {data/stitching-mpit-1sh.dat};
        \addlegendentry{\acs{uPIT} $T_\text{tr}=4\,s$}
        \addplot[mPIT,mark=o] table[x=stitcher_size, y=mpit8s] {data/stitching-mpit-1sh.dat};
        \addlegendentry{\acs{mPIT} $T_\text{tr}=8\,s$}
        \addplot[uPIT,mark=o] table[x=stitcher_size, y=upit8s] {data/stitching-mpit-1sh.dat};
        \addlegendentry{\acs{uPIT} $T_\text{tr}=8\,s$}
        \addplot[mPIT,mark=diamond] table[x=stitcher_size, y=mpit16s] {data/stitching-mpit-1sh.dat};
        \addlegendentry{\acs{mPIT} $T_\text{tr}=16\,s$}
        \addplot[uPIT,mark=diamond] table[x=stitcher_size, y=upit16s] {data/stitching-mpit-1sh.dat};
        \addlegendentry{\acs{uPIT} $T_\text{tr}=16\,s$}
        \addplot[mPIT,mark=star] table[x=stitcher_size, y=mpit32s] {data/stitching-mpit-1sh.dat};
        \addlegendentry{\acs{mPIT} $T_\text{tr}=32\,s$}



        \nextgroupplot[
legend cell align={left},
legend style={fill opacity=0.8, draw opacity=1, text opacity=1, draw=white!80!black},
xminorgrids,
ymin=0, ymax=1,
xticklabels={1, 2, 6, 14},
xlabel=$\currentcontext$ in s,
ylabel=\%,
ytick={0,0.5,1},
yticklabels={0, 50, 100},
y label style={yshift=-0.1cm},
]
        \path [draw=white, fill=palette-10]
(axis cs:1,0.00689655172413793)
--(axis cs:1,0)
--(axis cs:2,0)
--(axis cs:6,0)
--(axis cs:14,0)
--(axis cs:18,0)
--(axis cs:18,0)
--(axis cs:18,0)
--(axis cs:14,0)
--(axis cs:6,0)
--(axis cs:2,0)
--(axis cs:1,0.00689655172413793)
--cycle;

\path [draw=white, fill=palette-1]
(axis cs:1,0.46151724137931)
--(axis cs:1,0.00689655172413793)
--(axis cs:2,0)
--(axis cs:6,0)
--(axis cs:14,0)
--(axis cs:18,0)
--(axis cs:18,0)
--(axis cs:18,0)
--(axis cs:14,0.0813008130081301)
--(axis cs:6,0.188135593220339)
--(axis cs:2,0.370083102493075)
--(axis cs:1,0.46151724137931)
--cycle;

\path [draw=white, fill=palette-2]
(axis cs:1,0.969931034482759)
--(axis cs:1,0.46151724137931)
--(axis cs:2,0.370083102493075)
--(axis cs:6,0.188135593220339)
--(axis cs:14,0.0813008130081301)
--(axis cs:18,0)
--(axis cs:18,0)
--(axis cs:18,0)
--(axis cs:14,0.199186991869919)
--(axis cs:6,0.606779661016949)
--(axis cs:2,0.919667590027701)
--(axis cs:1,0.969931034482759)
--cycle;

\path [draw=white, fill=palette-3]
(axis cs:1,1)
--(axis cs:1,0.969931034482759)
--(axis cs:2,0.919667590027701)
--(axis cs:6,0.606779661016949)
--(axis cs:14,0.199186991869919)
--(axis cs:18,0)
--(axis cs:18,0)
--(axis cs:18,0)
--(axis cs:14,0.552845528455285)
--(axis cs:6,0.964406779661017)
--(axis cs:2,0.999445983379502)
--(axis cs:1,1)
--cycle;

\path [draw=white, fill=palette-4]
(axis cs:1,1)
--(axis cs:1,1)
--(axis cs:2,0.999445983379502)
--(axis cs:6,0.964406779661017)
--(axis cs:14,0.552845528455285)
--(axis cs:18,0)
--(axis cs:18,0)
--(axis cs:18,0)
--(axis cs:14,0.91869918699187)
--(axis cs:6,1)
--(axis cs:2,1)
--(axis cs:1,1)
--cycle;

\path [draw=white, fill=palette-5]
(axis cs:1,1)
--(axis cs:1,1)
--(axis cs:2,1)
--(axis cs:6,1)
--(axis cs:14,0.91869918699187)
--(axis cs:18,0)
--(axis cs:18,0.258064516129032)
--(axis cs:18,0.258064516129032)
--(axis cs:14,0.995934959349594)
--(axis cs:6,1)
--(axis cs:2,1)
--(axis cs:1,1)
--cycle;

\path [draw=white, fill=palette-6]
(axis cs:1,1)
--(axis cs:1,1)
--(axis cs:2,1)
--(axis cs:6,1)
--(axis cs:14,0.995934959349594)
--(axis cs:18,0.258064516129032)
--(axis cs:18,0.516129032258065)
--(axis cs:18,0.516129032258065)
--(axis cs:14,1)
--(axis cs:6,1)
--(axis cs:2,1)
--(axis cs:1,1)
--cycle;

\path [draw=white, fill=white!47.4509803921569!black]
(axis cs:1,1)
--(axis cs:1,1)
--(axis cs:2,1)
--(axis cs:6,1)
--(axis cs:14,1)
--(axis cs:18,0.516129032258065)
--(axis cs:18,0.806451612903226)
--(axis cs:18,0.806451612903226)
--(axis cs:14,1)
--(axis cs:6,1)
--(axis cs:2,1)
--(axis cs:1,1)
--cycle;

\path [draw=white, fill=palette-7]
(axis cs:1,1)
--(axis cs:1,1)
--(axis cs:2,1)
--(axis cs:6,1)
--(axis cs:14,1)
--(axis cs:18,0.806451612903226)
--(axis cs:18,1)
--(axis cs:18,1)
--(axis cs:14,1)
--(axis cs:6,1)
--(axis cs:2,1)
--(axis cs:1,1)
--cycle;

\plot[red,very thick] coordinates {
    (1, 0.97) (2,0.92) (6, 0.61) (14, 0.2) (18, 0)
};

\node[white] at (axis cs: 2,0.1) {1};
\node[white] at (axis cs: 4,0.6) {2};
\node[white] at (axis cs: 7,0.8) {3};
\node[white] at (axis cs: 12.5,0.8) {4};
\node[white] at (axis cs: 17.7,0.2) {5};
\node[white] at (axis cs: 17.7,0.45) {6};
\node[white] at (axis cs: 17.7,0.7) {7};
\node[white] at (axis cs: 17.7,0.94) {8};
\coordinate (c) at (axis cs: 18,0) ;
    \end{groupplot}
    \node[anchor=south,yshift=-0.1cm] at (group c1r1.north) {\ref{legend}};
    \node[inner sep=0,anchor=north,xshift=-1mm] at (c) {\tab{no\\stitching}};
\end{tikzpicture}